\documentclass[smus]{snow2e}
\def\MPL #1 #2 #3 {Mod.~Phys.~Lett.~{\bf#1},\ #2 (#3)}
\def\NPB #1 #2 #3 {Nucl.~Phys.~{\bf#1},\ #2 (#3)}
\def\PLB #1 #2 #3 {Phys.~Lett.~{\bf#1},\ #2 (#3)}
\def\PR #1 #2 #3 {Phys.~Rep.~{\bf#1},\ #2 (#3)}
\def\PRD #1 #2 #3 {Phys.~Rev.~{\bf#1},\ #2 (#3)}
\def\PRL #1 #2 #3 {Phys.~Rev.~Lett.~{\bf#1},\ #2 (#3)}
\def\RMP #1 #2 #3 {Rev.~Mod.~Phys.~{\bf#1},\ #2 (#3)}
\def\ZP #1 #2 #3 {Z.~Phys.~{\bf#1},\ #2 (#3)}
\def\IJMP #1 #2 #3 {Int.~J.~Mod.~Phys.~{\bf#1},\ #2 (#3)}
\def\lam{\lambda}
\def\lamu{\lam_u}
\def\lamd{\lam_d}
\def\lamud{\lam_u^\dagger}
\def\lamdd{\lam_d^\dagger}
\def\mgut{M_{\rm GUT}}

\def\mth{m_{3/2}}
\def\delgs{\delta_{GS}}
\def\kpr{K^\prime}
\def\eL{\tilde e_L}
\def\eR{\tilde e_R}

\def\sur{{\tilde{u}_R}}
\def\msur{{m_{\sur}}}
\def\ibid{{\it ibid.}}
\def\anti{\overline}
\def\call{{\cal L}}

\def\del{\delta}

\def\rts{\sqrt s}

\def\lam{\lambda}

\def\epem{e^+e^-}
\def\mupmum{\mu^+\mu^-}

\def\lsim{\mathrel{\raise.3ex\hbox{$<$\kern-.75em\lower1ex\hbox{$\sim$}}}}
\def\gsim{\mathrel{\raise.3ex\hbox{$>$\kern-.75em\lower1ex\hbox{$\sim$}}}}
\def\@versim#1#2{\vcenter{\offinterlineskip
 \ialign{$\m@th#1\hfil##\hfil$\crcr#2\crcr\sim\crcr } }}

\def\etmiss{\slash E_T}

\def\gam{\gamma}

\def\anti{\overline}

\def\fbi{~{\rm fb}^{-1}}
\def\fb{~{\rm fb}}

\def\GeV{\,{\rm GeV}}
\def\gev{\,{\rm GeV}}
\def\tev{\,{\rm TeV}}

\def\gl{{\tilde g}}
\def\mgl{m_{\gl}}
\def\stop{{\tilde t}}
\def\stopone{\stop_1}
\def\stoptwo{\stop_2}

\def\mstopone{m_{\stopone}}
\def\mstoptwo{m_{\stoptwo}}
\def\sbot{{\tilde b}}
\def\sbotone{\sbot_1}

\def\msbotone{m_{\sbotone}}

\def\stl{{\tilde{t}_L}}
\def\str{{\tilde{t}_R}}
\def\mstl{m_{\stl}}
\def\mstr{m_{\str}}
\def\sbl{{\tilde{b}_L}}
\def\sbr{{\tilde{b}_R}}
\def\msbl{m_{\sbl}}
\def\msbr{m_{\sbr}}

\def\slepl{\tilde{\ell}_L}
\def\mslepl{m_{\slepl}}
\def\slepr{\tilde{\ell}_R}
\def\mslepr{m_{\slepr}}

\def\hl{h^0}

\def\mhl{m_{\hl}}

\def\tanb{\tan\beta}

\def\mt{m_t}

\def\mz{m_Z}

\def\cnone{\tilde{\chi}^0_1}
\def\cntwo{\tilde{\chi}^0_2}

\def\mcnone{m_{\cnone}}
\def\mcntwo{m_{\cntwo}}

\def\cpone{\tilde{\chi}^+_1}
\def\cmone{\tilde{\chi}^-_1}
\def\cpmone{\tilde{\chi}^{\pm}_1}

\def\mcpmone{m_{\cpmone}}

\def\cpmtwo{\tilde{\chi}^{\pm}_2}

\def\mcpmtwo{m_{\cpmtwo}}

\input psfig.sty
\begin{document}
\title{Report of the Supersymmetry Theory Subgroup}

\author{
J.  Amundson (Wisconsin), G.  Anderson (FNAL), H.  Baer$^*$ (FSU),
J.  Bagger (Johns Hopkins),\\
R.M.  Barnett (LBNL), C.H.  Chen (UC  Davis), G.  Cleaver (OSU),
B.  Dobrescu (BU),\\
 M.  Drees (Wisconsin),
J.F.  Gunion (UC  Davis), G.L.  Kane (Michigan), B.  Kayser (NSF), \\
C.  Kolda (IAS), J.  Lykken\thanks{Theory subgroup conveners.} (FNAL),
S.P.  Martin (Michigan), T.  Moroi (LBNL), \\
S.  Mrenna (Argonne),
M.  Nojiri (KEK), D.  Pierce (SLAC), X.  Tata (Hawaii), \\
S.  Thomas (SLAC), J.D.  Wells (SLAC),
B.  Wright (North Carolina), Y.  Yamada (Wisconsin)
}

\maketitle

\thispagestyle{empty}\pagestyle{empty}
\begin{abstract}
We provide a mini-guide to some of the possible manifestations of
weak-scale supersymmetry.  For each of six scenarios we
provide
\begin{itemize}
\item a brief description of the theoretical underpinnings,
\item the adjustable parameters,
\item a qualitative description of the associated phenomenology at
future colliders,
\item comments on how to simulate each scenario
 with existing event generators.
\end{itemize}
\end{abstract}

\section{Introduction}

The Standard Model (SM) is a theory of spin-$1\over 2$ matter fermions
which interact via the exchange of spin-1 gauge bosons, where the bosons
and fermions live in independent representations of the gauge symmetries.
Supersymmetry (SUSY) is a symmetry which
establishes a one-to-one correspondence between bosonic and fermionic
degrees of freedom, and provides a relation between their
couplings \cite{bagger}.  Relativistic quantum field
theory is formulated to
be consistent with the symmetries of the Lorentz/Poincar\'e group -- a
non-compact Lie algebra.  Mathematically, supersymmetry is formulated
as a generalization of the Lorentz/Poincar\'e group of space-time
symmetries to include
spinorial generators which obey specific anti-commutation relations; such
an algebra is known as a graded Lie algebra.  Representations of the SUSY
algebra include both bosonic and fermionic degrees of freedom.

The hypothesis that nature is supersymmetric is very compelling to many
particle physicists for several reasons.
\begin{itemize}
\item It can be shown that the SUSY algebra is the only non-trivial
extension of the set of spacetime symmetries which forms one of the
foundations of relativistic quantum field theory.
\item If supersymmetry is formulated as a {\it local} symmetry, then one is
necessarily forced into introducing a massless spin-2 (graviton) field into
the theory. The resulting supergravity theory reduces to Einstein's
general relativity theory in the appropriate limit.
\item Spacetime supersymmetry appears to be a fundamental ingredient
of superstring theory.
\end{itemize}
These motivations say nothing about the {\it scale} at which nature might be
supersymmetric.  Indeed, there are additional motivations for
{\it weak-scale supersymmetry}.
\begin{itemize}
\item Incorporation of supersymmetry into the SM leads to a solution of the
gauge hierarchy problem.  Namely, quadratic divergences in loop
corrections to the Higgs boson mass will cancel between fermionic and bosonic
loops.  This mechanism works only if the superpartner particle masses are
roughly of order or less than the weak scale.
\item There exists an experimental hint: the three gauge couplings
can unify at the
Grand Unification scale if there exist weak-scale supersymmetric particles,
with a desert between the weak scale and the GUT scale.
This is not the case with the SM.
\item Electroweak symmetry breaking is a derived consequence of
supersymmetry breaking in many particle physics models with weak-scale
supersymmetry, whereas electroweak symmetry breaking in the SM is put in
``by hand.''  The SUSY radiative electroweak symmetry-breaking
mechanism works best if the top quark has mass
$m_t\sim 150-200$ GeV.  The recent discovery of the top quark with
$m_t=176\pm 4.4$ GeV is consistent with this mechanism.
\item As a bonus, many particle physics models with weak-scale
supersymmetry contain
an excellent candidate for cold dark matter (CDM): the lightest neutralino.
Such a CDM particle seems necessary to describe many aspects of cosmology.
\end{itemize}
Finally, there is a historical precedent for supersymmetry.  In 1928,
P.  A.  M.  Dirac incorporated the symmetries of the Lorentz group into
quantum mechanics.  He found as a natural consequence that each known
particle had to have a partner particle -- namely, antimatter.
The matter-anti-matter symmetry wasn't revealed until high enough
energy scales were reached to create a positron.  In a similar manner,
incorporation of supersymmetry into particle physics once again predicts
partner particles for all known particles.  Will nature prove to be
supersymmetric at the weak scale? In this report, we try to
shed light on some of the many possible ways that weak-scale supersymmetry
might be revealed by colliders operating at sufficiently high energy.

\subsection{Minimal Supersymmetric Standard Model}

The simplest supersymmetric model of particle physics which is consistent
with the SM is called
the Minimal Supersymmetric Standard Model (MSSM).  The recipe for this
model is to start with the SM of particle physics, but in addition
add an extra Higgs doublet of opposite hypercharge.  (This ensures
cancellation of triangle anomalies due to Higgsino partner contributions.)
Next, proceed with supersymmetrization, following well-known rules to
construct supersymmetric gauge theories.  At this stage one has a globally
supersymmetric SM theory.  Supersymmetry breaking is incorporated by adding
to the Lagrangian explicit soft SUSY-breaking terms consistent with the
symmetries of the SM.
These consist of scalar and gaugino
mass terms, as well as trilinear ($A$ terms) and bilinear ($B$ term) scalar
interactions.  The resulting theory has $>100$ parameters, mainly from the
various soft SUSY-breaking terms.
Such a model is the most conservative approach to
realistic SUSY model building, but the large parameter space leaves little
predictivity.  What is needed as well is a theory of how the soft
SUSY-breaking terms arise.  The fundamental field content of the MSSM is
listed in Table 1,
for one generation of quark and lepton (squark and slepton) fields.
Mixings and symmetry breaking lead to the actual physical mass eigenstates.

\begin{table}
\begin{center}
\begin{tabular}{c|cc}
\hline
\ & Boson fields & Fermionic partners \cr
\hline
Gauge multiplets & & \cr
$SU(3)$ & $g^a$ & $\tilde g^a$ \cr
$SU(2)$ & $W^i$ & $\tilde{W}^i$ \cr
$U(1)$ & $B$ & $\tilde B$ \cr
\hline
Matter multiplets & & \cr
Scalar leptons & $\tilde{L}^j=(\tilde{\nu} ,\tilde{e}^-_L)$ & $(\nu
,e^-)_L$\cr
\ & $\tilde{R}=\tilde{e}^+_R$ & $e_L^c$ \cr
Scalar quarks & $\tilde{Q}^j=(\tilde{u}_L,\tilde{d}_L)$ & $(u,d)_L$ \cr
\ & $\tilde{U}=\tilde{u}_R^*$ & $u_L^c$ \cr
\ & $\tilde{D}=\tilde{d}_R^*$ & $d_L^c$ \cr
Higgs bosons & $H_1^j$ & $(\tilde{H}_1^0,\tilde{H}_1^-)_L$ \cr
\ & $H_2^j$ & $(\tilde{H}_2^+,\tilde{H}_2^0)_L$ \cr
\hline
\end{tabular}
\caption{Field content of the MSSM for one generation of
quarks and leptons.}
\label{mssm}
\end{center}
\end{table}

The goal of this report is to create a mini-guide to some of the possible
supersymmetric models that occur in the literature, and to provide a
bridge between SUSY model builders and their experimental colleagues.  The
following sections each contain a brief survey of six classes of
SUSY-breaking models studied at this workshop;
contributing group members are listed in {\it italics}.
We start with the most popular framework for
experimental searches, the paradigm
\begin{itemize}
\item minimal supergravity model (mSUGRA) ({\it M.  Drees and M.  Nojiri}),
\end{itemize}
and follow with
\begin{itemize}
\item models with additional D-term contributions to scalar masses,
({\it C.  Kolda, S.  Martin and S.  Mrenna})
\item models with non-universal GUT-scale soft SUSY-breaking terms,
({\it G.  Anderson, R.  M.  Barnett, C.  H.  Chen, J.  Gunion, J.  Lykken, T.
Moroi
and Y.  Yamada})
\item two MSSM scenarios which use the large parameter freedom of
the MSSM to fit to various collider zoo events, ({\it G.  Kane and S.  Mrenna})
\item models with $R$ parity violation, ({\it H.  Baer, B.  Kayser and X.
Tata})
 and
\item models with gauge-mediated low energy SUSY breaking (GMLESB),
({\it J.  Amundson, C.  Kolda, S.  Martin, T.  Moroi, S.  Mrenna, D.  Pierce,
S.  Thomas, J.  Wells and B.  Wright}).
\end{itemize}

Each section contains a brief description of the model, qualitative
discussion of some of the associated phenomenology, and finally some
comments on event generation for the model under discussion.  In this way,
it is hoped that this report will be a starting point for future
experimental SUSY searches, and that it will provide a flavor for the
diversity of ways that weak-scale supersymmetry might manifest itself
at colliding beam experiments.  We note that a survey of some additional
models is contained in Ref.  \cite{peskin}, although under a somewhat
different format.

\section{Minimal Supergravity Model}

The currently most popular SUSY model is the minimal supergravity
(mSUGRA) model \cite{sug1,sug2}.  Here one assumes that SUSY is broken
spontaneously in a ``hidden sector,'' so that some auxiliary field(s)
get vev(s) of order $M_Z \cdot M_{Pl} \simeq (10^{10} \ {\rm
GeV})^2$.  Gravitational -- strength interactions then {\em automatically}
transmit SUSY breaking to the ``visible sector,'' which
contains all the SM fields and their superpartners; the effective mass
splitting in the visible sector is by construction of order of the
weak scale, as needed to stabilize the gauge hierarchy.  In
{\em minimal} supergravity one further assumes that the kinetic terms for the
gauge and matter fields take the canonical form: as a result,
all scalar fields
(sfermions and Higgs bosons) get the same contribution $m_0^2$ to
their squared scalar masses, and that all trilinear $A$ parameters
have the same value $A_0$, by virtue of an approximate
global $U(n)$ symmetry of
the SUGRA Lagrangian \cite{sug2}.
Finally, motivated by the apparent
unification of the measured gauge couplings within the MSSM
\cite{sug3} at scale $\mgut \simeq 2 \cdot 10^{16}$ GeV, one assumes
that SUSY-breaking gaugino masses have a common value $m_{1/2}$
at scale $\mgut$.  In practice, since little is known about physics
between the scales $\mgut$ and $M_{\rm Planck}$,
one often uses $\mgut$ as the scale
at which the scalar masses and $A$ parameters unify.  We note that
$R$ parity is assumed to be conserved within the mSUGRA framework.

This ansatz has several advantages.  First, it is very economical; the
entire spectrum can be described with a small number of free
parameters.  Second, degeneracy of scalar masses at scale $\mgut$ leads
to small flavor-changing neutral currents.
Finally, this model predicts radiative breaking of the
electroweak gauge symmetry \cite{sug4} because of the large top-quark
mass.

Radiative symmetry breaking together with the precisely known value of
$M_Z$ allows one to trade two free parameters, usually taken to be the
absolute value of the supersymmetric Higgsino mass parameter $|\mu |$ and
the $B$ parameter appearing in the scalar Higgs potential, for the
ratio of vevs, $\tan \beta$.  The model then has four continuous and one
discrete free parameter not present in the SM:
\begin{equation}
m_0, m_{1/2}, A_0, \tan \beta, {\rm sign}(\mu).
\end{equation}

This model is now incorporated in several publicly available MC codes,
in particular {\tt ISAJET} \cite{isajet}.
An approximate version is incorporated into {\tt Spythia} \cite{spythia},
which reproduces {\tt ISAJET} results to 10\%.
Most SUSY spectra studied at this workshop have been
generated within mSUGRA; we refer to the various accelerator subgroup
reports for
the corresponding spectra.  One ``generically'' finds the following
features:
\begin{itemize}
\item $|\mu|$ is large, well above the masses of the $SU(2)$ and
$U(1)$ gauginos.  The lightest neutralino is therefore mostly a
Bino (and an excellent candidate for cosmological CDM -- for related
constraints, see {\it e.g.} Ref.  \cite{cosmo}),
and the second neutralino and lighter chargino are dominantly
$SU(2)$ gauginos.  The heavier neutralinos and
charginos are only rarely produced in the decays of gluinos and
sfermions (except possibly for stop decays).  Small regions of parameter
space with $|\mu | \simeq M_W$ are possible.
\item If $m_0^2 \gg m_{1/2}^2$, all sfermions of the first two
generations are close in mass.  Otherwise, squarks are significantly
heavier than sleptons, and $SU(2)$ doublet sleptons are heavier than singlet
sleptons.  Either way, the lighter stop and sbottom eigenstates are
well below the first generation squarks; gluinos therefore have
large branching ratios into $b$ or $t$ quarks.
\item The heavier Higgs bosons (pseudoscalar $A$, scalar $H^0$,
and charged $H^\pm$) are usually heavier than $|\mu|$ unless
$\tan\beta \gg 1$.  This also implies that the light scalar $h^0$
behaves like the SM Higgs.
\end{itemize}

These features have already become something like folklore.  We want to
emphasize here that even within this restrictive framework, quite
different spectra are also possible, as illustrated by the following
examples.

Example A is for $m_0 = 750$ GeV, $m_{1/2} = 150$ GeV, $A_0 = -300$
GeV, $\tan \beta = 5.5$, $\mu<0$, and $m_t=165$ GeV (pole mass).  This
yields $|\mu| = 120$ GeV, very similar to the $SU(2)$ gaugino mass
$M_2$ at the weak scale, leading to strong Higgsino -- gaugino
mixing.  The neutralino masses are 60, 91, 143 and 180 GeV, while
charginos are at 93 and 185 GeV.  They are all considerably lighter
than the gluino (at 435 GeV), which in turn lies well below the
squarks (at $\simeq$ 815 GeV) and sleptons (at 750-760 GeV).  Due to
the strong gaugino -- Higgsino mixing, all chargino and neutralino
states will be produced with significant rates in the decays of
gluinos and $SU(2)$ doublet sfermions, leading to complicated decay
chains.  For example, the $\ell^+ \ell^-$ invariant mass spectrum in gluino
pair events will have many thresholds due to $\tilde{\chi}^0_i
\rightarrow \tilde{\chi}^0_j \ell^+\ell^-$ decays.  Since first and second
generation squarks are almost twice as heavy as the gluino, there
might be a significant gluino ``background'' to squark production at
the LHC.  A 500 GeV $e^+e^-$ collider will produce all six chargino and
neutralino states.  Information about $\tilde{e}_L, \ \tilde{e}_R$ and
$\tilde{\nu}_e$ masses can be gleaned from studies of neutralino and
chargino production, respectively; however, $\sqrt{s}>$ 1.5 TeV is
required to study sleptons directly.  Spectra of this type can already
be modelled reliably using {\tt ISAJET}: the above parameter space set can be
entered via the $SUGRA$ keyword.

As example B, we have chosen $m_0=m_{1/2}=200$ GeV, $A_0=0$, $\tan
\beta =48$, $\mu < 0$ and $m_t=175$ GeV.  Note the large value of $\tan
\beta$, which leads to large $b$ and $\tau$ Yukawa couplings, as
required in models where all third generation Yukawa couplings are
unified at scale $\mgut$.  Here the gluino (at 517 GeV) lies slightly
above first generation squarks (at 480-500 GeV), which in turn lie
well above first generation sleptons (at 220-250 GeV).  The
light neutralinos (at 83 and 151 GeV) and light chargino (at 151 GeV)
are mostly gauginos, while the heavy states (at 287, 304 and
307 GeV) are mostly Higgsinos, because $\vert\mu\vert=275 \
{\rm GeV}\gg m_{1/2}$.

The masses of $\tilde{t}_1$ (355 GeV), $\tilde{b}_1$ (371 GeV) and
$\tilde{\tau}_1$ (132 GeV) are all significantly below those of the
corresponding first or second generation sfermions.  As a result, more
than 2/3 of all gluinos decay into a $b$ quark and a $\tilde{b}$
squark.  Since (s)bottoms have large Yukawa couplings, $\tilde{b}$
decays will often produce the heavier, Higgsino-like chargino and
neutralinos.  Further, all neutralinos (except for the lightest one,
which is the LSP) have two-body decays into $\tilde{\tau}_1 + \tau$;
in case of $\tilde{\chi}^0_2$ this is the only two-body mode, and for
the Higgsino-like states this mode will be enhanced by the large
$\tau$ Yukawa coupling.  Chargino decays will also often produce real
$\tilde{\tau}_1$.  Study of the $\ell^+\ell^-$ invariant mass spectrum will
not allow direct determination of neutralino mass differences, as the
$\ell^\pm$ are secondaries from tau decays.  Even $\tilde{e}_L$ pair
events at $e^+e^-$ colliders will contain up to four tau leptons!
Further, unless the $e^-$ beam is almost purely right-handed, it
might be difficult to distinguish between $\tilde{\tau}_1$ pair
production and $\tilde{\chi}_1^\pm$ pair production.  Finally, the
heavier Higgs bosons are quite light in this case, e.g.  $m_A = 126$
GeV.  There will be a large number of $A \rightarrow
\tau^+ \tau^-$ events at the LHC.  However, because most SUSY events
will contain $\tau$ pairs in this scenario, it is not clear whether
the Higgs signal will remain visible.  At present, scenarios with
$\tan \beta \gg 1$ can not be simulated with {\tt ISAJET}, since the $b$
and $\tau$ Yukawa couplings have not been included in all relevant
decays.  This situation should be remedied soon.


\section{$D$-term Contributions to Scalar Masses}

We have seen that the standard mSUGRA framework predicts
a testable pattern of squark and slepton masses.
In this section we
describe a class of models in which a quite distinctive modification
of the mSUGRA predictions can arise, namely contributions
to scalar masses associated with the
$D$-terms of extra spontaneously broken gauge symmetries
 \cite{dterm}.
As we will see, the modification of squark, slepton and Higgs masses can
have a profound effect on phenomenology.

In general, $D$-term contributions to scalar masses will arise in
supersymmetric models whenever a gauge symmetry is spontaneously
broken with a reduction of rank.
Suppose,
for example, that the SM gauge group $SU(3)\times SU(2)\times U(1)_Y$
is supplemented by an additional $U(1)_X$ factor broken far above
the electroweak scale.  Naively, one might suppose that if the
breaking scale is sufficiently large, all direct effects of
$U(1)_X$ on TeV-scale physics are negligible.  However, a simple
toy model shows that this is not so.
Assume that ordinary MSSM scalar fields, denoted generically by
$\varphi_i$, carry $U(1)_X$ charges $X_i$ which are not all 0.
In order to break $U(1)_X$,
we also assume the existence of a pair of additional
chiral superfields $\Phi$ and $\overline \Phi$ which are SM singlets,
but carry $U(1)_X$ charges
which are normalized (without loss of generality) to be $+1$ and $-1$
respectively.
Then VEV's for $\Phi$ and $\overline\Phi$ will spontaneously
break $U(1)_X$ while leaving the SM gauge group intact.
The scalar potential whose minimum determines
$\langle\Phi\rangle,\langle\overline\Phi\rangle$
then has the form
\begin{equation}
 V=V_{0}
+ m^2 |\Phi|^2 + {\overline m}^2 |{\overline \Phi}|^2
+ {g_X^2\over 2 } \left [ |\Phi|^2 - |{\overline \Phi}|^2
+ X_i |\varphi_i |^2 \right ]^2.
\end{equation}
Here $V_0$ comes from the superpotential
and involves only $\Phi$ and $\overline\Phi$;
it is symmetric under $\Phi\leftrightarrow\overline\Phi$, but otherwise its
precise form need not concern us.
The pieces involving $m^2$ and ${\overline m}^2$
are soft breaking terms; $m^2$ and ${\overline m}^2$ are
of order $M_Z^2$ and in general unequal.
The remaining piece is the square of the
$D$-term associated
with $U(1)_X$, which forces the minimum of the potential to occur along
a nearly $D$-flat direction
$\langle \Phi \rangle\approx \langle \overline\Phi \rangle $.
This scale can be much larger than 1 TeV with
natural choices of $V_0$, so that the $U(1)_X$ gauge boson
is very heavy and plays no role in collider physics.

However, there is also a deviation from $D$-flatness given by
$\langle \Phi \rangle^2
- \langle \overline\Phi \rangle^2
 \approx D_X/g_X^2$, with $D_X =({\overline m}^2 - m^2)/2$, which directly
affects the
masses of the remaining light MSSM fields.  After integrating out $\Phi$
and $\overline\Phi$, one finds that each MSSM scalar (mass)$^2$
receives a correction given by
\begin{equation}
\Delta m_i^2 = X_i D_X
\label{dtermcorrections}
\end{equation}
where $D_X$ is again typically of order
$M_Z^2$
and may have either sign.
This result does not depend on the scale at which $U(1)_X$ breaks; this
turns out to be a general feature,
independent of assumptions about the precise mechanism
of symmetry breaking.
Thus $U(1)_X$ manages to leave its ``fingerprint" on the masses of the
squarks,
sleptons, and Higgs bosons, even if it is broken at an arbitrarily high
energy.  From a TeV-scale point of view, the parameter $D_X$ might
as well be taken as a parameter of our ignorance regarding physics
at very high energies.  The important point is that $D_X$ is universal,
so that each MSSM scalar (mass)$^2$ obtains a contribution
simply proportional
to $X_i$, its charge under $U(1)_X$.
Typically the $X_i$ are
rational numbers and do not all have the same sign, so that a particular
candidate $U(1)_X$
can leave a quite distinctive pattern of mass splittings on the squark and
slepton spectrum.

The extra $U(1)_X$ in this discussion may stand alone, or may be
embedded in a larger non-abelian gauge group,
perhaps together with the SM
gauge group (for example in an
$SO(10)$ or $E_6$ GUT).
If the gauge group contains more than one $U(1)$ in addition to $U(1)_Y$, then
each $U(1)$ factor can contribute a set of corrections exactly analogous to
(\ref{dtermcorrections}).  Additional $U(1)$ groups are endemic
in superstring models, so at least from that point of view one
may be optimistic about the existence of corresponding $D$-terms
and their potential importance in the study of the squark and slepton
mass spectrum at future colliders.
It should be noted that once one assumes the existence of additional
gauged $U(1)$'s at very high energies, it is quite unnatural to
assume that $D$-term contributions to scalar masses can be avoided
altogether.  (This would require an exact symmetry
enforcing $m^2 = {\overline m}^2$ in the example above.)
The only question is whether or not
the magnitude of the $D$-term contributions
is significant compared to the usual mSUGRA contributions.
Note also that as long as the charges $X_i$ are
family-independent, then from (\ref{dtermcorrections})
squarks and sleptons
with the same electroweak quantum numbers remain degenerate,
maintaining the natural suppression of flavor changing neutral currents.

It is not difficult to implement the effects of $D$-terms in simulations,
by imposing the corrections (\ref{dtermcorrections}) to a particular
``template" mSUGRA model.
After choosing the $U(1)_X$ charges of the MSSM fields, our remaining
ignorance of the mechanism of $U(1)_X$ breaking is
parameterized by $D_X$ (roughly of order $M_Z^2$).
The $\Delta m_i^2$ corrections should be imposed at the scale
$M_X$ where one chooses to assume that $U(1)_X$ breaks.
(If $M_X < M_{\rm Planck}$ or $M_{\rm GUT}$,
one should also in principle incorporate
renormalization group effects due to $U(1)_X$ above $M_X$,
but these can often be shown to be small.)
The other parameters of the theory are
unaffected.
One can then run these parameters down to the electroweak
scale, in exactly the same way as in mSUGRA models, to find the
spectrum of sparticle masses.

(The solved-for parameter $\mu$ is then indirectly affected by $D$-terms,
through the requirement of correct electroweak symmetry breaking.)
The only subtlety involved is an apparent ambiguity in choosing the
charges $X_i$, since any linear combination of $U(1)_X$ and $U(1)_Y$
charges might be used.
These charges should be picked to correspond to the basis
in which there is no mixing in the kinetic
terms of the $U(1)$ gauge bosons.  In particular models where $U(1)_X$
and/or $U(1)_Y$ are embedded in non-abelian groups, this linear
combination is uniquely determined; otherwise it can be arbitrary.

A test case which seems particularly worthy of study is that of
an additional gauged $B-L$ symmetry.  In this case the $U(1)_X$ charges
for each MSSM scalar field are a linear combination of $B-L$ and $Y$.
If this model is embedded in $SO(10)$ (or certain of its subgroups),
then the unmixed linear combination of $U(1)$'s appropriate
for (\ref{dtermcorrections}) is
$X = -{5\over 3}(B-L) + {4\over 3} Y$.  The $X$ charges for the
MSSM squarks and sleptons are $-1/3$ for $Q_L,u_R,e_R$ and $+1$
for $L_L$ and
$d_R$.  The MSSM Higgs fields have charges $+2/3$ for $H_u$ and
$-2/3$ for $H_d$.  Here we consider the modifications to a
mSUGRA model defined by the parameters
$(m_0,m_{1/2},A_0)=(200,100,0)$~GeV, $\mu<0$, and $\tan\beta=2$,
assuming $m_t=175$ GeV.

The effects of $D$-term contributions to the scalar mass spectrum is
illustrated in Fig.~1, which shows the masses of
$\tilde{e}_L,\tilde{e}_R$, the lightest Higgs boson $h$, and the
lightest bottom squark $\tilde{b}_1$ as a function of $D_X$.
The unmodified mSUGRA prediction is found at
$D_X=0$.  A particularly dramatic
possibility is that $D$-terms could invert the usual hierarchy of slepton
masses, so that $m_{{\tilde e}_L},m_{{\tilde \nu}} < m_{{\tilde e}_R}$.
In the test model, this occurs for negative $D_X$; the negative endpoint
of $D_X$ is set by the experimental lower bound on $m_{\tilde \nu}$.
The relative change of the squark masses is smaller, while the change
to the lightest Higgs boson mass is almost negligible except near
the positive $D_X$ endpoint where it reaches the experimental lower
bound.  The complicated mass spectrum perhaps can be probed most
directly at the NLC with precision measurements of squark and
slepton masses.
Since the usual MSSM renormalization group contributions to scalar
masses are much larger for squarks than for sleptons, it is likely
that the effects of $D$-term contributions are relatively
larger for sleptons.

At the Tevatron and LHC, it has been suggested in these proceedings that
SUSY parameter determinations can be obtained by making global fits
of the mSUGRA parameter space to various observed signals.  In this
regard it should be noted that significant $D$-term contributions
could invalidate such strategies unless they are generalized.
This is because adding
$D$-terms (\ref{dtermcorrections}) to a given template mSUGRA model
can dramatically change certain branching fractions
by altering the kinematics of decays involving squarks and especially
sleptons.  This is demonstrated for the test model in Fig.~2.
Thus we find for example that the product
$BR({\tilde\chi}^+_1 \rightarrow \ell^+X)\times
BR({\tilde\chi}_2^0 \rightarrow \ell^+\ell^- X)$ can change
up to an order of magnitude or more as one varies $D$-terms
(with all other parameters held fixed).
Note that the branching ratios of Fig.~2 include the leptons
from two-body and three-body decays, e.g.  ${\tilde\chi}^+_1 \rightarrow
\ell^+\nu {\tilde\chi}_1^0$
and ${\tilde\chi}^+_1 \rightarrow {\tilde \ell}^+\nu
\rightarrow \ell^+ {\tilde\chi}_j^0 \nu$.
On the other hand,
the $BR({\tilde g}\rightarrow b X)$ is fairly insensitive to
$D$-terms over most, but not all, of parameter space.

Since the squark masses are generally much less affected by
the $D$-terms, and the gluino mass only indirectly,
the production cross sections for squarks and gluinos
should be fairly stable.  Therefore, the variation of
$BR({\tilde g}\rightarrow b X)$ is an accurate gauge of
the variation of observables such as the $b$ multiplicity
of SUSY events.  Likewise, the ${\tilde\chi}^\pm_1 {\tilde\chi}_2^0$
production cross section does not change much as the
$D$-terms are varied, so the
expected trilepton signal can vary like the product
of branching ratios -- by orders of magnitude.
While the results presented are for a specific,
and particularly simple, test model,
similar variations can be observed in other
explicit models.
The possible presence of $D$-terms should be considered when
interpreting a SUSY signal at future colliders.  An
experimental analysis
which proves or disproves their existence would be a unique
insight into physics at very high energy scales.

To facilitate event generation, approximate expressions for the modified
mass spectra are implemented in the {\tt Spythia} Monte Carlo,
assuming the $D$-terms are added in at the unification scale.
Sparticle spectra from models with extra $D$-terms can be incorporated
into {\tt ISAJET} simply via the $MSSMi$ keywords, although the user must
supply a program to generate the relevant spectra via RGE's or
analytic formulae.

\begin{figure}[h]
\hskip 1cm\psfig{figure=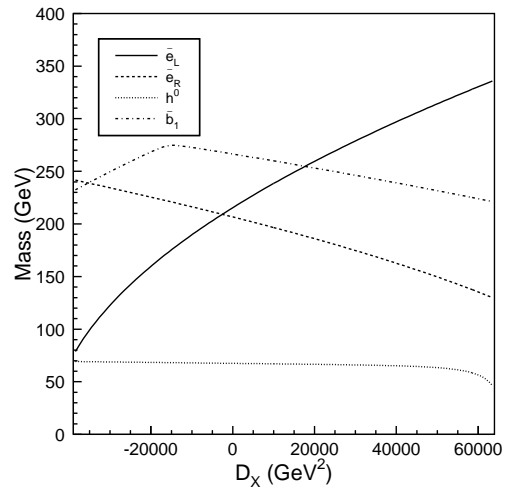,width=7cm}
\caption{Mass spectrum as a function of $D_{X}$.}
\end{figure}
\begin{figure}[h]
\hskip 1cm\psfig{figure=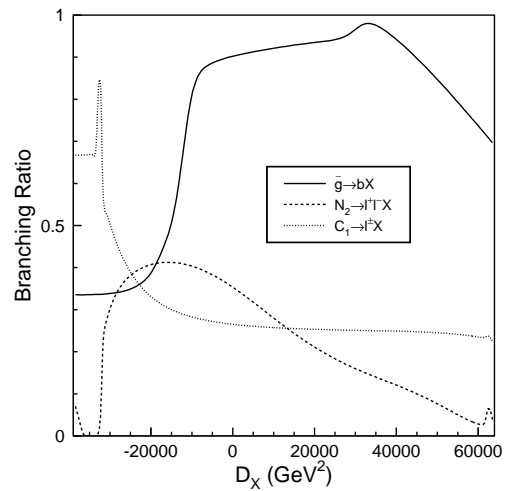,width=7cm}
\caption{Branching ratios as a function of $D_{X}$.}
\end{figure}


\section{Non-Universal GUT-Scale Soft SUSY-Breaking Parameters}

\subsection{Introduction}

We considered models in which the gaugino masses and/or the
scalar masses are not universal at the GUT scale, $\mgut$.
We study the extent to which non-universal boundary conditions
can influence experimental signatures and
detector requirements,
and the degree to which experimental data can distinguish
between different models for the
GUT-scale boundary conditions.

\subsubsection{Non-Universal Gaugino Masses at $\mgut$}

We focus on two well-motivated types of models:

\noindent $\bullet$ Superstring-motivated models
in which SUSY breaking is
moduli dominated.  We consider
the particularly attractive O-II model of Ref.~ \cite{Ibanez}.
The boundary conditions at $\mgut$ are:
\begin{equation}
\begin{array}{l}
M_a^0\sim \sqrt 3 \mth[-(b_a+\delgs)K\eta] \\
m_0^2=\mth^2[-\delgs\kpr] \\
A_0=0
\end{array}
\label{bcs}
\end{equation}
where $b_a$ are SM beta function coefficients, $\delgs$
is a mixing parameter,
which would be a negative integer in the O-II model,
and $\eta=\pm1$.
{}From the estimates of Ref.~ \cite{Ibanez},
$K \simeq 4.6\times 10^{-4}$ and $\kpr \simeq 10^{-3}$, we
expect that slepton and squark masses would be very much
larger than gaugino masses.

\noindent $\bullet$ Models
in which SUSY breaking occurs via an $F$-term that is not
an $SU(5)$ singlet.  In this class of models, gaugino masses are generated
by a chiral superfield $\Phi$ that appears linearly in the gauge
kinetic function, and whose auxiliary $F$ component acquires an
intermediate-scale vev:
\begin{equation}
\call\sim \int d^2\theta W^a W^b {\Phi_{ab}\over M_{\rm Planck}} + h.c.
\sim {\langle F_{\Phi} \rangle_{ab}\over M_{\rm Planck}}
\lam^a\lam^b\, +\ldots ,
\end{equation}
where the $\lam^{a,b}$ are the gaugino fields.
$F_{\Phi}$
belongs to an $SU(5)$ irreducible representation which
appears in the symmetric product of two adjoints:
\begin{equation}
({\bf 24}{\bf \times}
{\bf 24})_{\rm symmetric}={\bf 1}\oplus {\bf 24} \oplus {\bf 75}
 \oplus {\bf 200}\,,
\label{irrreps}
\end{equation}
where only $\bf 1$ yields universal masses.
Only the component of $F_{\Phi}$ that is ```neutral'' with respect to
the SM gauge group should acquire a vev,
$\langle F_{\Phi} \rangle_{ab}=c_a\del_{ab}$, with $c_a$
then determining the relative magnitude of
the gauginos masses at $\mgut$: see Table~\ref{masses}.

\begin{table}
\begin{center}
\begin{small}
\begin{tabular}{|c|ccc|ccc|}
\hline
\ & \multicolumn{3}{c|} {$\mgut$} & \multicolumn{3}{c|}{$\mz$} \cr
$F_{\Phi}$
& $M_3$ & $M_2$ & $M_1$
& $M_3$ & $M_2$ & $M_1$ \cr
\hline
${\bf 1}$ & $1$ &$\;\; 1$ &$\;\;1$ & $\sim \;6$ & $\sim \;\;2$ &
$\sim \;\;1$ \cr
${\bf 24}$ & $2$ &$-3$ & $-1$ & $\sim 12$ & $\sim -6$ &
$\sim -1$ \cr
${\bf 75}$ & $1$ & $\;\;3$ &$-5$ & $\sim \;6$ & $\sim \;\;6$ &
$\sim -5$ \cr
${\bf 200}$ & $1$ & $\;\; 2$ & $\;10$ & $\sim \;6$ & $\sim \;\;4$ &
$\sim \;10$ \cr
\hline
$\stackrel{\textstyle O-II}{\delgs=-4}$ & $1$ & $\;\;5$ & ${53\over 5}$ &
$\sim 6$ & $\sim 10$ & $\sim {53\over5}$ \cr
\hline
\end{tabular}
\end{small}
\caption{Relative gaugino masses at $\mgut$ and $\mz$
in the four possible $F_{\Phi}$ irreducible representations,
and in the O-II model with $\delgs\sim -4$.}
\label{masses}
\end{center}
\end{table}

\begin{figure}[htb]
\leavevmode
\centerline{\psfig{file=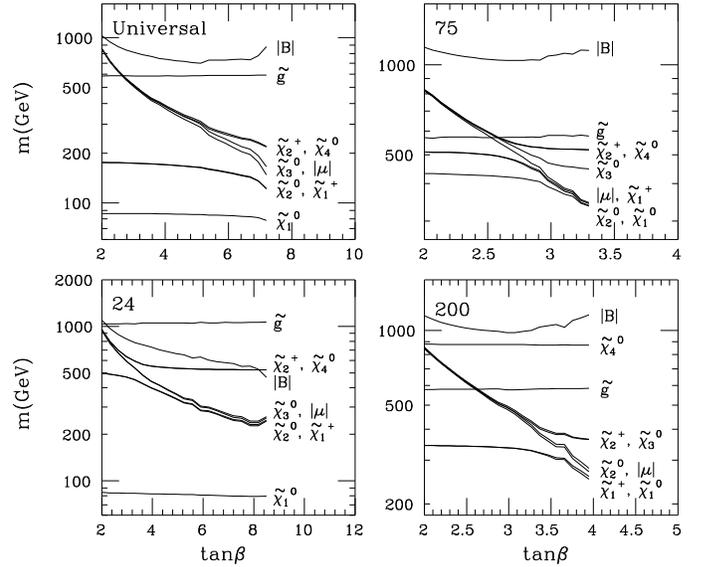,width=3.5in}}
\caption{Physical (pole) gaugino masses as a function of $\tanb$
for the ${\bf 1}$ (universal), ${\bf 24}$, ${\bf 75}$, and ${\bf 200}$
$F$ representation choices.  Also plotted are $|B|$ and $|\mu|$.
We have taken $m_0=1\tev$ and $M_3=200,400,200,200\gev$,
respectively.}
\label{mtanb}
\end{figure}

Physical masses of the gauginos are
influenced by $\tanb$-dependent off-diagonal terms in the mass
matrices and by corrections which boost $\mgl(pole)$ relative to $\mgl(\mgl)$.
If $\mu$ is large, the lightest neutralino (which is the LSP)
will have mass $\mcnone\sim {\rm min}(M_1,M_2)$ while the lightest
chargino will have $\mcpmone\sim M_2$.  Thus, in the ${\bf 200}$
and O-II scenarios with
$M_2\lsim M_1$, $\mcpmone\simeq\mcnone$ and the $\cpmone$
and $\cnone$ are both Wino-like.  The
$\tanb$ dependence of the masses at $\mz$ for the universal,
${\bf 24}$, ${\bf 75}$, and ${\bf 200}$ choices appears in Fig.~\ref{mtanb}.
The $\mgl$-$\mcnone$ mass splitting becomes increasingly
smaller in the sequence ${\bf 24}$, ${\bf 1}$, ${\bf 200}$
${\bf 75}$, O-II,
as could be anticipated from Table~\ref{masses}.
It is interesting to note that at
high $\tanb$, $\mu$ decreases to a level comparable to $M_1$ and
$M_2$, and there is substantial degeneracy among the $\cpmone$, $\cntwo$
and $\cnone$.

\subsubsection{Non-Universal Scalar Masses at $\mgut$}

We consider
models in which the SUSY-breaking scalar masses at $\mgut$ are influenced
by the Yukawa couplings of the corresponding quarks/leptons.
This idea is exemplified in the model of Ref.~ \cite{hallrandall}
based on perturbing about the $[U(3)]^5$ symmetry that is present
in the absence of Yukawa couplings.  One finds, for example:
\begin{equation}
{\bf m}_{\tilde Q}^2=m_0^2(I+c_Q\lamud\lamu+c_Q^\prime\lamdd\lamd+\ldots)
\end{equation}
where $Q$ represents the squark partners of the left-handed quark doublets.
The Yukawas $\lamu$ and $\lamd$ are $3\times 3$ matrices
in generation space.
The $\ldots$ represent terms of order $\lam^4$ that we will neglect.
A priori, $c_Q$, $c_Q^\prime$, should all be similar in size,
in which case the large top-quark Yukawa coupling implies that
the primary deviations from universality will occur in $\mstl^2$,
$\msbl^2$ (equally and in the same direction).\footnote{In this discussion
we neglect an analogous, but independent, shift in $\mstr^2$.}
It is the fact that $\mstl^2$ and $\msbl^2$ are shifted equally
that will distinguish $m^2$ non-universality from the effects of a large
$A_0$ parameter at $\mgut$; the latter would primarily
introduce $\stl-\str$ mixing and yield a low
$\mstopone$ compared to $\msbotone$.

\subsection{Phenomenology}

\subsubsection{Non-Universal Gaugino Masses}

We examined the phenomenological
implications for the standard Snowmass comparison point
({\it e.g.} NLC point \#3) specified by $\mt=175\gev$, $\alpha_s=0.12$,
$m_0=200\gev$, $M_3^0=100\gev$, $\tanb=2$, $A_0=0$ and $\mu$$<$$0$.
In treating the O-II model
we take $m_0=600\gev$, a value that yields a (pole)
value of $\mgl$ not unlike that for the other scenarios.
The masses of the supersymmetric particles for
each scenario are given in Table~\ref{susymasses}.

\begin{table}
\begin{center}
\begin{small}
\begin{tabular}{|c|c|c|c|c|c|}
\hline
\ & ${\bf 1}$ & ${\bf 24}$ & ${\bf 75}$ & ${\bf 200}$ &

$\stackrel{\textstyle O-II}{\delgs=-4.7}$ \cr
\hline

$\mgl$ & 285 & 285 & 287 & 288 & 313 \cr
$\msur$ & 302 & 301 & 326 & 394 & - \cr
$\mstopone$ & 255 & 257 & 235 & 292 & - \cr
$\mstoptwo$ & 315 & 321 & 351 & 325 & - \cr
$\msbl$ & 266 & 276 & 307 & 264 & - \cr
$\msbr$ & 303 & 303 & 309 & 328 & - \cr
$\mslepr$ & 207 & 204 & 280 & 437 & - \cr
$\mslepl$ & 216 & 229 & 305 & 313 & - \cr
$\mcnone$ & 44.5 & 12.2 & 189 & 174.17 & 303.09 \cr
$\mcntwo$ & 97.0 & 93.6 & 235 & 298 & 337 \cr
$\mcpmone$ & 96.4 & 90.0 & 240 & 174.57 & 303.33 \cr
$\mcpmtwo$ & 275 & 283 & 291 & 311 & - \cr
$\mhl$ & 67 & 67 & 68 & 70 & 82 \cr
\hline
\end{tabular}
\end{small}
\caption{Sparticle masses for the Snowmass comparison point
in the different gaugino mass scenarios.  Blank entries for the O-II
model indicate very large masses.}
\label{susymasses}
\end{center}
\end{table}

The phenomenology of these scenarios for $\epem$ collisions
is not absolutely straightforward.

\begin{itemize}
\item
In the $\bf 75$ model, $\cpone\cmone$ and $\cntwo\cntwo$
pair production at $\rts=500\gev$ are
barely allowed kinematically; the phase space for $\cnone\cntwo$
is only somewhat better.  All the signals would be rather weak,
but could probably be extracted with sufficient integrated luminosity.

\item
In the $\bf 200$ model, $\epem\to \cpone\cmone$ production would be
kinematically allowed at a $\rts=500\gev$ NLC, but not easily observed
due to the fact that the (invisible) $\cnone$ would take essentially all of
the energy in the $\cpmone$ decays.  However,
according to the results of Ref.~ \cite{moddominated},
$\epem\to \gam\cpone\cmone$ would be observable at $\rts=500\gev$.

\item
The O-II model with $\delgs$ near $-4$ predicts that $\mcpmone$
and $\mcnone$ are both rather close to $\mgl$, so that
$\epem\to \cpone\cmone,\cnone\cnone$ would {\it not} be kinematically allowed
at  $\sqrt s = 500\gev$.
The only SUSY ``signal'' would be the presence of a very SM-like
light Higgs boson.
\end{itemize}

At the LHC, the strongest signal for SUSY would arise from $\gl\gl$
production.  The different models lead to very distinct signatures
for such events.  To see this, it is sufficient to list the primary
easily identifiable decay chains of the gluino for each of the five scenarios.
(In what follows, $q$ denotes any quark other than a $b$.)
\begin{eqnarray*}
 {\bf 1:} && \gl \stackrel{90\%}{\to} \sbl\anti b \stackrel{99\%}{\to}
\cntwo b\anti b\stackrel{33\%}{\to} \cnone(\epem~{\rm or}~\mupmum)b\anti b\\
 \phantom{{\bf 1:}} &&
\phantom{\gl \stackrel{90\%}{\to} \sbl\anti b \stackrel{99\%}{\to}
\cntwo b\anti b}
\stackrel{8\%}{\to} \cnone \nu\anti\nu b\anti b\\
 \phantom{{\bf 1:}} &&
\phantom{\gl \stackrel{90\%}{\to} \sbl\anti b \stackrel{99\%}{\to}
\cntwo b\anti b}
\stackrel{38\%}{\to} \cnone q\anti q b\anti b\\
 \phantom{{\bf 1:}} &&
\phantom{\gl \stackrel{90\%}{\to} \sbl\anti b \stackrel{99\%}{\to}
\cntwo b\anti b}
\stackrel{8\%}{\to} \cnone b\anti b  b\anti b\\
 {\bf 24:} && \gl \stackrel{85\%}{\to} \sbl\anti b
\stackrel{70\%}{\to} \cntwo b\anti b\stackrel{99\%}{\to}
 \hl \cnone b\anti b\stackrel{28\%}{\to} \cnone b\anti b b\anti b \\
\phantom{ {\bf 24:}} &&
\phantom{ \gl \stackrel{85\%}{\to} \sbl\anti b
\stackrel{70\%}{\to} \cntwo b\anti b\stackrel{99\%}{\to}
 \hl \cnone b\anti b}
\stackrel{69\%}{\to} \cnone \cnone\cnone b\anti b \\
 {\bf 75:} && \gl \stackrel{43\%}{\to} \cnone g~{\rm or}~\cnone q\anti q \\
\phantom{{\bf 75:}} && \phantom{\gl}\stackrel{10\%}{\to} \cnone b\anti b \\
\phantom{{\bf 75:}} && \phantom{\gl}\stackrel{20\%}{\to}
\cntwo g~{\rm or}~\cntwo q\anti q \\
\phantom{{\bf 75:}} && \phantom{\gl}\stackrel{10\%}{\to} \cntwo b\anti b \\
\phantom{{\bf 75:}} && \phantom{\gl}\stackrel{17\%}{\to} \cpmone q\anti q \\
 {\bf 200:} && \gl \stackrel{99\%}{\to} \sbl\anti b
\stackrel{100\%}{\to} \cnone b\anti b \\
 {\bf O-II{\bf:}} && \gl \stackrel{51\%}{\to} \cpmone q\anti q \\
\phantom{{\rm O-II{\bf:}}} && \phantom{\gl} \stackrel{17\%}{\to} \cnone g \\
\phantom{{\rm O-II{\bf:}}} &&
\phantom{\gl} \stackrel{26\%}{\to} \cnone q\anti q \\
\phantom{{\rm O-II{\bf:}}} &&
\phantom{\gl} \stackrel{6\%}{\to} \cnone b\anti b
\end{eqnarray*}

Gluino pair production will then lead to the following strikingly
different signals.

\begin{itemize}
\item
In the $\bf 1$ scenario we expect a very large
number of final states with missing energy,
four $b$-jets and two lepton-antilepton pairs.

\item
For $\bf 24$, an even larger number of events will have missing energy
and eight $b$-jets, four of which reconstruct to two pairs with
mass equal to (the known) $\mhl$.

\item
The signal for $\gl\gl$ production in the case of $\bf 75$ is
much more traditional; the primary decays yield
multiple jets (some of which are
$b$-jets) plus $\cnone$, $\cntwo$ or $\cpmone$.
Additional jets, leptons and/or neutrinos
arise when $\cntwo\to\cnone$ + two jets,
two leptons or two neutrinos or $\cpmone\to\cnone$ +
two jets or lepton+neutrino.

\item
In the $\bf 200$ scenario, we find missing energy plus four $b$-jets;
only $b$-jets appear in the primary decay -- any other
jets present would have to come from initial- or final-state radiation,
and would be expected to be softer on average.  This is almost
as distinctive a signal as the $8b$ final state found in the $\bf 24$
scenario.

\item
In the final O-II scenario, $\cpmone\to \cnone$ + very soft
spectator jets or leptons that would not be easily detected.  Even
the $q\anti q$ or $g$ from the primary decay would not be very energetic
given the small mass splitting between $\mgl$ and $\mcpmone\sim\mcnone$.
Soft jet cuts would have to be used to dig out this signal,
but it should be possible given the very high $\gl\gl$ production rate
expected for this low $\mgl$ value; see Ref.~ \cite{moddominated}.
\end{itemize}

Thus, for the Snowmass comparison point,
distinguishing between the different boundary condition scenarios
at the LHC will be easy.  Further, the event rate
for a gluino mass this low is such that the end-points of
the various lepton, jet or $\hl$ spectra will allow relatively good
determinations of the mass differences between the sparticles appearing
at various points in the final state decay chain.
We are optimistic that this will prove to be
a general result so long as event rates are large.

\subsubsection{Non-Universal Scalar Masses}

Once again we focus on the Snowmass overlap point.  We maintain
gaugino mass universality at $\mgut$, but allow for non-universality
for the squark masses.  Of the many possibilities, we focus
on the case where only $c_Q\neq 0$ with $A_0=0$ (as assumed
for the Snowmass overlap point).  The phenomenology for this case
is compared to that which would emerge if we take $A_0\neq 0$
with all the $c_i=0$.

Consider the $\gl$ branching ratios as a function
of $\mstl$$=$$\msbl$ as $c_Q$ is varied from negative to positive values.
As the common mass crosses the threshold above which
the $\gl\to \sbotone b$ decay becomes kinematically disallowed,
we revert to a more standard SUSY scenario in which $\gl$
decays are dominated by modes such as $\cpmone q\anti q$,
$\cnone q\anti q$, $\cntwo q\anti q$ and $\cntwo b\anti b$.
For low enough $\mstl$, the $\gl\to \stopone t$ mode opens
up, but must compete with the $\gl\to\sbotone b$ mode
that has even larger phase space.

In contrast, if $A_t$ is varied,
the $\gl$ branching ratios remain essentially constant until
$\mstopone$ is small enough that $\gl\to \stopone t$ is kinematically
allowed.  Below this point, this latter mode quickly dominates
the $\sbotone b$ mode which continues to have very small phase
space given that the $\sbotone$ mass remains essentially constant
as $A_t$ is varied.

\subsection{Event Generation}

A thorough search and determination
of the rates (or lack thereof) for the full panoply of possible
channels is required to distinguish the many possible
GUT-scale boundary conditions from one another.  In the program {\tt ISAJET},
independent weak-scale gaugino masses may be input using the $MSSM4$
keyword.  Independent third generation
squark masses may be input via the $MSSM2$ keyword.
The user must supply a program to generate the relevant weak-scale parameter
values from the specific GUT-scale assumptions.
Relevant weak-scale MSSM parameters can also be input to {\tt Spythia};
as with {\tt ISAJET}, the user must provide a program for the specific
model.

\newcommand{ \PLBold }[3]{Phys.  Lett.  {\bf #1B} (#2) #3}
\newcommand{ \PRold }[3]{Phys.  Rev.  {\bf #1} (#2) #3}
\newcommand{ \PREP }[3]{Phys.  Rep.  {\bf #1} (#2) #3}
\newcommand{ \ZPC }[3]{Z.  Phys.  C {\bf #1} (#2) #3}
\def\slashchar#1{\setbox0=\hbox{$#1$} 
 \dimen0=\wd0 
 \setbox1=\hbox{/} \dimen1=\wd1 
 \ifdim\dimen0>\dimen1 
 \rlap{\hbox to \dimen0{\hfil/\hfil}} 
 #1 
 \else 
 \rlap{\hbox to \dimen1{\hfil$#1$\hfil}} 
 / 
 \fi} %

\section{MSSM Scenarios Motivated by Data}

An alternative procedure for gleaning
information about the SUSY soft terms is to use the full
(> 100 parameters) parameter space freedom of the MSSM and match
to data, assuming one has a supersymmetry signal.  This approach has been
used in the following two examples.

\subsection{The CDF $e^+e^-\gamma\gamma +\etmiss$ Event}

Recently a candidate for sparticle production has been reported
 \cite{event} by the CDF collaboration.
This has been interpreted in several
ways \cite{PRL}, \cite{DimopoulosPRL}, \cite{DimopoulosSecond},
 \cite{Grav} and later with additional variations \cite{kolda},
 \cite{LopezNanopoulos}, \cite{Hisano}.  The main two paths are whether
the LSP is the lightest neutralino \cite{PRL}, \cite{sandro}, or a
nearly massless gravitino \cite{DimopoulosPRL,DimopoulosSecond,
Grav,kolda,LopezNanopoulos} or axino \cite{Hisano}.  In the
gravitino or axino case the LSP is not a candidate for cold dark matter,
SUSY can have no effect on $R_b$ or $\alpha^Z_s$ or $BR(b\to s\gamma),$
and stops and gluinos are not being observed at FNAL.  In the case where
the lightest neutralino is the LSP, the opposite holds for all of
these observables, and we will pursue this case in detail here.

The SUSY Lagrangian depends on a number of parameters, all of
which have the dimension of mass.  That should not be viewed as a weakness
because at present we have no theory of the origin of mass parameters.
Probably getting such a theory will depend on understanding how
SUSY is broken.  When there is no data on sparticle masses and
couplings, it is appropriate to make simplifying assumptions,
based on theoretical prejudice,
to reduce the number of parameters.  However, once there may be data,
it is important to constrain the most general set of parameters and
see what patterns emerge.
We proceed by making no assumptions about soft breaking parameters.
In practice, even though the full theory has over a hundred such
parameters, that is seldom a problem since any given observable depends
on at most a few.

The CDF event \cite{event} has a 36 GeV $e^-$, a 59 GeV $e^+$, photons of
38 and 30 GeV, and $\slashchar{E}_T = $ 53 GeV.
A SUSY interpretation is $q\bar q\to \gamma^{*}, Z^{*} \to {\tilde e}^+
{\tilde e}^-$, followed by each ${\tilde e}^\pm \to e^\pm \tilde{\chi}_2^0,$
$\tilde{\chi}_2^0 \to \gamma\tilde{\chi}_1^0.$ The second lightest
neutralino, $\tilde{\chi}_2^0$, must be
photino-like since it couples strongly to $\tilde ee$.  Then the
LSP = $\tilde{\chi}_1^0$ must be
Higgsino-like \cite{Komatsu,HaberWyler,AmbrosMele1} to
have a large $BR(\tilde{\chi}_2^0 \to\tilde{\chi}_1^0 \gamma).$
The range of parameter choices for
this scenario are given in Table \ref{eeggtab}.

\begin{table}
\begin{center}
\begin{tabular}{|c|c|} \hline
\multicolumn{2}{|c|}{$e^+e^-\gamma\gamma + \etmiss$
constraints on supersymmetric parameters}
 \\ \hline \hline
$\eL$ & $\eR$ \\ \hline
$100 \lsim m_{\eL} \lsim 130 \; \GeV$
 & $100 \lsim m_{\eR} \lsim 112 \; \GeV$ \\
$50 \lsim M_1 \lsim 92 \; \GeV$
 & $60 \lsim M_1 \lsim 85 \; \GeV$ \\
$50 \lsim M_2 \lsim 105 \; \GeV$
 & $40 \lsim M_2 \lsim 85 \; \GeV$ \\
$0.75 \lsim M_2/M_1 \lsim 1.6$
 & $0.6 \lsim M_2/M_1 \lsim 1.15$ \\
$-65 \lsim \mu \lsim -35 \; \GeV$
 & $-60 \lsim \mu \lsim -35 \; \GeV$ \\
$0.5 \lsim |\mu|/M_1 \lsim 0.95$
 & $0.5 \lsim |\mu|/M_1 \lsim 0.8$ \\
$1 \lsim \tan \beta \lsim 3 $
 & $1 \lsim \tan \beta \lsim 2.2$ \\ \hline
\end{tabular}
\caption{Constraints on the MSSM parameters and masses in the
neutralino LSP scenario.
}
\label{eeggtab}
\end{center}
\end{table}

If light superpartners indeed exist, FNAL and LEP will
produce thousands of them, and measure their properties very well.
The first thing to check at FNAL is whether the produced selectron is
${\tilde e}_L$ or ${\tilde e}_R.$ If ${\tilde e}_L,$ then the charged current
channel $u\overline{d} \to W^+ \to {\tilde e}_L \tilde \nu$ has 5-10 times
the rate of ${\tilde e}_L^+ {\tilde e}_L^-$.
We expect ${\tilde e}_L \to e\tilde{\chi}_2^0(\to \gamma\tilde{\chi}_1^0).$
Most likely
 \cite{sandro} $\tilde \nu \to e \tilde{\chi}_1^{\pm},$
where $\tilde{\chi}_1^{\pm}$ is the lightest
chargino.  If the stop mass $m_{\tilde t} < m_{\tilde{\chi}_1^{\pm}}$, then
$\tilde{\chi}_1^{\pm} \to \tilde t (\to c\tilde{\chi}_1^0)b$
so $\tilde \nu \to ebc\tilde{\chi}_1^0$; if
$m_{\tilde t} > m_{\tilde{\chi}_1^{\pm}}$
then $\tilde{\chi}_1^{\pm} \to W^* (\to jj)\tilde{\chi}_1^0$ so $\tilde \nu
\to ejj\tilde{\chi}_1^0,$ where $j=u,d,s,c.$
Either way, dominantly ${\tilde e}_L \tilde \nu
\to ee\gamma jj \slashchar{E}_T$ where $j$ may be light or heavy quarks.
If no such signal is found, probably the produced selectron was ${\tilde e}_R$.
Also, $\sigma (\tilde\nu\tilde\nu) \cong \sigma({\tilde e}_L {\tilde e}_L)$.
Cross sections for many channels are given in Ref.~ \cite{sandro}.

The most interesting channel (in our opinion)
at FNAL is $u\overline{d} \to W^+ \to \tilde{\chi}^+_i
\tilde{\chi}_2^0.$ This gives a signature $\gamma jj \slashchar{E}_T,$ for
which
there is only small parton-level SM backgrounds.
If $m_{\tilde t} < m_{\tilde{\chi}^{\pm}_i}$, one of $j$
is a $b.$ If $t\to \tilde t \tilde{\chi}_2^0$ (expected about 10\%
of the time) and,
if $\tilde q$ are produced at FNAL, there are additional sources of such
events (see below).


If charginos, neutralinos and sleptons are light, then gluinos and
squarks may not be too heavy.  If stops are light ($m_{{\tilde t}_1}\simeq
M_W$), then $BR(t\to
\tilde t \tilde{\chi}^0_i) \simeq 1/2$ \cite{wells}.
In this case, an extra source
of tops must exist beyond SM production,
because $\sigma \times BR(t\to Wb)^2$ is near or above its SM
value with $BR(t\to Wb)=1.$
With these motivations,
the authors of \cite{KaneMrenna}
have suggested that one assume $m_{\tilde g} \geq m_t + m_{\tilde t}$
and $m_{\tilde q} \geq m_{\tilde g}$, with
$m_{\tilde q}\simeq 250-300$ GeV.  Then there are several pb of top
production via channels $\tilde q \tilde g, \tilde g \tilde g, \tilde q
\bar{\tilde q}$ with $\tilde q \to q \tilde g,$ and $\tilde g \to t
\tilde t$ since $t\tilde t$ is the gluino's only two-body
decay mode.
This analysis points out that
$P_T(t\bar t)$ should peak at smaller $P_T$ for the SM than
for the SUSY scenario, since the system is recoiling against extra jets in
the SUSY case.
The SUSY case suggests that if
$m_t$ or $\sigma_{t\bar t}$ are measured in different channels one will
obtain different values, which may be consistent with reported data.
This analysis also argues that the present data is consistent with
$BR(t\to \tilde t \tilde{\chi}^0_i)= 1/2.$

At present \cite{Rapporteur} $R_b$ and $BR(b\to s\gamma)$ differ from their
SM predictions by 1.5-2$\sigma$, and $\alpha_s$ measured by the $Z$
width differs by about 1.5-2$\sigma$ from its value measured in DIS and
other ways.  If these effects are real they can be explained by
$\tilde{\chi}^{\pm}_i$ - $\tilde t$ loops,
using the same SUSY parameters
deduced from the $ee\gamma\gamma$ event ($+$ a light, mainly
right-handed, stop).  Although $\tan\beta, \mu,$ and $M_2$ a priori could be
anything, they come out the same from the analysis of these loops as
from $ee\gamma\gamma$ ($\tan\beta \leq 1.5, \mu \sim -m_Z/2, M_2 \sim
60-80$ GeV).

The LSP=$\tilde{\chi}_1^0$ apparently escapes the CDF detector in the
$ee\gamma\gamma$ event, suggesting it is stable (though only proving it
lives longer than $\sim 10^{-8}$ sec).  If so it is a candidate for CDM.
The properties of $\tilde{\chi}_1^0$ are deduced from the
analysis \cite{sandro} so the calculation of the relic density
 \cite{KaneWells} is highly constrained.  The analysis shows that the
s-channel annihilation of $\tilde{\chi}_1^0\tilde{\chi}_1^0$ through the $Z$
dominates, so the needed
parameters are $\tan\beta$, $m_{\tilde{\chi}_1^0}$
and the Higgsino fraction for
$\tilde{\chi}_1^0$, which is large.
The results are encouraging, giving $0.1
\leq \Omega h^2 \leq 1,$ with a central value $\Omega h^2 \simeq 1/4.$

The parameter choices of Table \ref{eeggtab} can be
input to event generators such as {\tt Spythia} or {\tt ISAJET}
(via $MSSMi$ keywords) to check that
the event rate and kinematics of the $ee\gamma\gamma$ event are
satisfied and then to determine other related signatures.  {\tt Spythia}
includes the $\tilde\chi_2^0\rightarrow\tilde{\chi}_1^0\gamma$ branching
ratio for low $\tan\beta$ values; for {\tt ISAJET}, the
$\tilde{\chi}_2^0\rightarrow\tilde{\chi}_1^0\gamma$ branching must be input
using the $FORCE$ command, or must be explicitly added into the decay table.

%

\subsection{CDF/D0 Dilepton Plus Jets Events}

Recently, CDF and D0 have reported various dilepton plus multi-jet
events which are presumably top-quark candidate events.  For several of
these events, however, the
event kinematics do not match well with those expected from a top
quark with mass $m_t\sim 175$ GeV.  The authors of Ref.  \cite{barnett}
have shown that the match to event kinematics can be improved by
hypothesizing a supersymmetry source for the recalcitrant events.
The supersymmetry source is best matched by considering
$\tilde q\tilde q$ production, where each
$\tilde q\rightarrow q\tilde{\chi},\tilde{\chi}\rightarrow\nu\tilde{\ell},
\tilde{\ell}\rightarrow\ell\tilde{\chi}_1^0$.  A recommended set of parameters
is as follows \cite{barnett}:
$m_{\tilde g}\simeq 330$ GeV, $m_{\tilde q}\simeq 310$ GeV,
$m_{\tilde{\ell}_L}\simeq 220$ GeV, $m_{\tilde\nu}\simeq 220$ GeV,
$m_{\tilde{\ell}_R}\simeq 130$ GeV, $\mu\simeq -400$ GeV,
$M_1\simeq 50$ GeV and $M_2\simeq 260$ GeV.  Note that this parameter
set discards the common hypothesis of gaugino mass unification.
These parameters can be input into {\tt Spythia} or
{\tt ISAJET} (via $MSSMi$ keywords), taking care to use the non-unified
gaugino masses as inputs.

\section{$R$ Parity Violation}

\def\topfraction{1.}
\def\textfraction{0.}

\def\gtrsim{\mathrel{\rlap{\raise 1pt \hbox{$>$}}}
 {\rlap{\lower 2pt \hbox{$\sim$}}}
 }
\def\sgn{\mathop{\rm sgn}}
\def\etmiss{\slashchar{E}_T}
\def\fb{{\rm fb}}
\def\fbi{{\rm fb}^{-1}}
\def\Meff{M_{\rm eff}}
\def\Msusy{M_{\rm SUSY}}
\def\lsp{\tilde{\chi}_1^0}
\def\ra{\rightarrow}
\def\GeV{{\rm GeV}}
\def\TeV{{\rm TeV}}

\let\badcite= \cite
\def \cite{~\badcite}

\def\slashchar#1{\setbox0=\hbox{$#1$} 
 \dimen0=\wd0 
 \setbox1=\hbox{/} \dimen1=\wd1 
 \ifdim\dimen0>\dimen1 
 \rlap{\hbox to \dimen0{\hfil/\hfil}} 
 #1 
 \else 
 \rlap{\hbox to \dimen1{\hfil$#1$\hfil}} 
 / 
 \fi} %

\def\dofig#1#2{\centerline{\epsfxsize=#1\epsfbox{#2}}%
 \vskip-.2in}

$R$ parity ($R$) is a quantum number which is +1 for any ordinary
particle, and -1 for any sparticle.  $R$-violating $(\slashchar{R})$
interactions occur naturally in supersymmetric theories, unless they
are explicitly forbidden.  Each $\slashchar{R}$ coupling also violates
either lepton number $L$, or baryon number $B$.  Together, these
couplings violate both $L$ and $B$, and lead to tree-level diagrams
which would make the proton decay at a rate in gross violation
of the observed bound.  To forbid such rapid decay, such
$\slashchar{R}$ couplings are normally set to zero.  However, what if such
couplings are actually present?

In supersymmetry with minimal field content, the allowable
$\slashchar{R}$ part of the superpotential is

\begin{equation}
W_{\slashchar{R}}=\lambda_{ijk}L_iL_j\bar E_k
 + \lambda^{\prime}_{ijk} L_iQ_j \bar D_k
 + \lambda^{\prime\prime}_{ijk} \bar U_i\bar D_j\bar D_k.
\end{equation}
Here, $L$, $Q$, $\bar E$, $\bar U$, and $\bar D$ are superfields
containing, respectively, lepton and quark doublets, and charged
lepton, up quark, and down quark singlets.  The indices $i,j,k$, over
which summation is implied, are generational indices.  The first term
in $W_{\slashchar{R}}$ leads to $L$-violating $(\slashchar{L})$
transitions such as $e+\nu_\mu\to\tilde e$.  The second one leads to
$\slashchar{L}$ transitions such as $u+\bar d\to\bar{\tilde e}$.  The
third one produces $\slashchar{B}$ transitions such as $\bar u+\bar
d\to\tilde d$.  To forbid rapid proton decay, it is often assumed that
if $\slashchar{R}$ transitions are indeed present, then only the
$L$-violating $\lambda$ and $\lambda^{\prime}$ terms occur, or only
the $B$-violating $\lambda^{\prime\prime}$ term occurs, but not both.
While the flavor components of $\lambda '\lambda ''$ involving $u,\ d,\ s$
are experimentally constrained to be $<10^{-24}$ from proton decay limits,
the other components of $\lambda '\lambda ''$ and $\lambda \lambda ''$ are
significantly less tightly constrained.

Upper bounds on the $\slashchar{R}$ couplings $\lambda$,
$\lambda^{\prime}$, and $\lambda^{\prime\prime}$ have been inferred
from a variety of low-energy processes, but most of these bounds are
not very stringent.  An exception is the bound on
$\lambda^{\prime}_{111}$, which comes from the impressive lower limit
of $9.6\times 10^{24}{\rm yr}$ \cite{1} on the half-life for the
neutrinoless double beta decay ${}^{76}{\rm Ge}\to {}^{76}{\rm
Se}+2e^-$.  At the quark level, this decay is the process $2d\to
2u+2e^-$.  If $\lambda^{\prime}_{111}\ne O$, this process can be
engendered by a diagram in which two $d$ quarks each undergo the
$\slashchar{R}$ transition $d\to\tilde u+e^-$, and then the two
produced $\tilde u$ squarks exchange a $\tilde g$ to become two $u$
quarks.  It can also be engendered by a diagram in which $2d\to
2\tilde d$ by $\tilde g$ exchange, and then each of the $\tilde d$
squarks undergoes the $\slashchar{R}$ transition
$\tilde d\to u+e^-$.  Both of these diagrams are proportional to
$\lambda^{\prime 2}_{111}$.  If we assume that the squark masses
occurring in the two diagrams are equal, $m_{{\tilde u}_{L}}\simeq
m_{{\tilde d}_{R}}\equiv m_{\tilde q}$, the previously quoted limit on
the half-life implies that \cite{2}

\begin{equation}
\vert\lambda^{\prime}_{111}\vert < 3.4\times 10^{-4}
 \left(\frac{m_{\tilde q}}{100\ \GeV}\right)^2
 \left(\frac{m_{\tilde g}}{100\ \GeV}\right)^{1/2}.
\end{equation}

It is interesting to recall that if the amplitude for neutrinoless
double beta decay is, for whatever reason, nonzero, then the electron
neutrino has a nonzero mass \cite{3}.  Thus, if
$\lambda^{\prime}_{1jj}\ne 0$, SUSY interactions lead to nonzero
neutrino mass \cite{3p5}.

The way \cite{4} in which low-energy processes constrain many of the
$\slashchar{L}$ couplings $\lambda$ and $\lambda^{\prime}$ is
illustrated by consideration of nuclear $\beta^-$ decay and $\mu^-$
decay.  In the Standard Model (SM), both of these decays result from
$W$ exchange alone, and the comparison of their rates tells us about
the CKM quark mixing matrix.  However, in the presence of
$\slashchar{R}$ couplings, nuclear $\beta^-$ decay can receive a
contribution from $\tilde d$, $\tilde s$, or $\tilde b$ exchange, and
$\mu^-$ decay from $\tilde e$, $\tilde \mu$, or $\tilde \tau$
exchange.  The information on the CKM elements which has been inferred
assuming that only $W$ exchange is present bounds these new
contributions, and it is found, for example, that \cite{4}
\begin{equation}
\vert \lambda_{12k}\vert < 0.04
\left(\frac{m_{{\tilde e}^k_R}}{100\ \GeV}\right),
\end{equation}
for each value of the generation index $k$.  In a similar fashion, a
number of low-energy processes together imply \cite{4} that for many of the
$\slashchar{L}$ couplings $\lambda_{ijk}$ and
$\lambda^{\prime}_{ijk}$,
\begin{equation}
\vert \lambda^{(\prime)}_{ijk}\vert < (0.03 \to 0.26)
 \left(\frac{m_{\tilde f}}{100\ \GeV}\right).
\end{equation}
Here, $m_{\tilde f}$ is the mass of the sfermion relevant to the bound
on the particular $\lambda^{(\prime)}_{ijk}$.

Bounds of order 0.1 have also been placed on the $\slashchar{L}$ couplings
$\lambda^{\prime}_{1jk}$ by searches for squarks formed through the action of
these couplings in $e^+p$ collisions at
HERA \cite{4h}.

Constraints on the $\slashchar{B}$ couplings $\lambda^{\prime\prime}$
come from nonleptonic weak processes which are suppressed in the SM,
such as rare $B$ decays and $K-\bar K$ and $D-\bar D$ mixing \cite{5}.
For example, the decay $B^+\to\overline{K^0}K^+$ is a penguin (loop)
process in the SM, but in the presence of $\slashchar{R}$ couplings
could arise from a tree-level diagram involving ${\tilde u}^k_R$
($k=1,2$, or $3$) exchange.  The present upper bound on the branching
ratio for this decay \cite{6} implies that \cite{5}
\begin{equation}
\vert \lambda^{\prime\prime}_{k12}\lambda^{\prime\prime}_{k23}\vert^{1/2}
 < 0.09 \left(\frac{m_{\tilde{u}^k_R}}{100\ \GeV}\right);\ k=1,2,3.
\end{equation}
Recently, bounds $\lambda'_{12k}<0.29$ and $\lambda'_{22k}<0.18$ for
$m_{\tilde q}=100$ GeV have been obtained from data on $D$ meson
decays \cite{3p5}.  For a recent review of constraints on $R$-violating
interactions, see Ref.  \cite{6p5}.

We see that if sfermion masses are assumed to be of order 100 GeV or
somewhat larger, then for many of the $\slashchar{R}$ couplings
$\lambda_{ijk}$,
$\lambda^{\prime}_{ijk}$ and $\lambda^{\prime\prime}_{ijk}$, the
existing upper bound is $\sim$ 0.1 for a sfermion mass of 100 GeV.
We note that this upper bound is comparable to
the values of some of the SM gauge
couplings.  Thus, $\slashchar{R}$ interactions could still prove to
play a significant role in high-energy collisions.

What effects of $\slashchar{R}$ might we see, and how would
$\slashchar{R}$ interactions affect future searches for SUSY? Let us
assume that $\slashchar{R}$ couplings are small enough that sparticle
production and decay are still dominated by gauge interactions, as in
the absence of $\slashchar{R}$.  The main effect of $\slashchar{R}$ is
then that the LSP is no longer
stable, but decays into ordinary particles, quite possibly within the
detector in which it is produced.  Thus, the LSP no longer carries
away transverse energy, and the missing transverse energy $(\etmiss)$
signal, which is the mainstay of searches for SUSY when $R$ is assumed
to be conserved, is greatly degraded.  (Production of SUSY particles
may still involve missing $E_T$, carried away by neutrinos.)

At future $e^+e^-$ colliders, sparticle production may include the processes
$e^+e^-\to\tilde{\chi}_i^+\tilde{\chi}_j^-$,
$\tilde{\chi}^0_i\tilde{\chi}^0_j$,
$\tilde{e}^+_L\tilde{e}^-_L$, $\tilde{e}^+_L\tilde{e}^-_R$,
$\tilde{e}^+_R \tilde{e}^-_L$, $\tilde{e}^+_R\tilde{e}^-_R$,
$\tilde{\mu}^+_L \tilde{\mu}^-_L$, $\tilde{\mu}^+_R\tilde{\mu}^-_R$,
$\tilde{\tau}^+_L\tilde{\tau}^-_L$, $\tilde{\tau}^+_R\tilde{\tau}^-_R$,
$\tilde{\nu}_L\bar{\tilde{\nu}}_L$.  Here, the $\tilde{\chi}^{\pm}_i$ are
charginos, and the $\tilde{\chi}^0_i$ are neutralinos.  Decay of the
produced sparticles will often yield high-$E_T$ charged leptons, which
can be sought in seeking evidence of SUSY.  Now, suppose the LSP is
the lightest neutralino, $\tilde{\chi}^0_1$.  If the $\slashchar{L}$,
$\slashchar{R}$ couplings $\lambda$ are nonzero, the $\tilde{\chi}^0_1$
can have the decays $\tilde{\chi}^0_1\to\mu\bar e\nu,\ e\bar e\nu$.

These yield high-energy leptons, so the strategy of looking for the
latter to seek evidence of SUSY will still work.  However, if the
$\slashchar{B}$, $\slashchar{R}$ couplings $\lambda^{\prime\prime}$
are nonzero, the $\tilde{\chi}^0_1$ can have the decays
$\tilde{\chi}^0_1\to cds, \bar c\bar d\bar s$.  When followed by these
decays, the production process $e^+e^-\to\tilde{\chi}^0_1\tilde{\chi}^0_1$
yields six jets which form a pair of three-jet systems.  The
invariant mass of each system is $m_{{\tilde{\chi}^0_1}}$, and there is
no missing energy.  This is quite an interesting signature.

Nonvanishing $\slashchar{L}$ and $\slashchar{R}$ couplings
$\lambda$ would also make possible resonant sneutrino production in $e^+e^-$
collisions. \cite{4} For example, we could have
$e^+e^-\to\tilde{\nu}_\mu\to\tilde{\chi}^{\pm}_1\mu^{\mp},
\tilde{\chi}^0_1\nu_\mu$.  At the resonance peak, the cross section
times branching ratio could be large \cite{4}.

In future experiments at hadron colliders, one can seek evidence of
gluino pair production by looking for the multilepton signal that may
result from cascade decays of the gluinos.  This signal will be
affected by the presence of $\slashchar{R}$ interactions.  The worst
case is where the LSP decays via $\slashchar{B}$, $\slashchar{R}$
couplings to yield hadrons.  The presence of these hadrons can cause
leptons in SUSY events to fail the lepton isolation criteria,
degrading the multilepton signal \cite{7}.  This reduces considerably the reach
in $m_{\tilde g}$ of the Tevatron.  At the
Tevatron with an integrated luminosity of 0.1 fb$^{-1}$, there is {\it no}
reach in $m_{\tilde g}$, while for 1 fb$^{-1}$
it is approximately 200 GeV \cite{7}, if $m_{\tilde q}=2m_{\tilde g}$.
At the LHC with an
integrated luminosity of 10 fb$^{-1}$, the reach extends beyond
$m_{\tilde g}=1\ \TeV$, even in the presence of $\slashchar{B}$ and
$\slashchar{R}$ interactions \cite{8}.

If $\slashchar{R}$ couplings are large, then conventional SUSY event
generators will need many production and decay mechanisms to be re-computed.
The results would be very model dependent, owing to the large parameter space
in the $\slashchar{R}$ sector.  If $\slashchar{R}$ couplings are assumed
small, so that gauge and Yukawa interactions still dominate production
and decay mechanisms, then event generators can be used by simply
adding in the appropriate expected decays of the LSP (see the approach in
Ref.  \cite{7,8}).  For {\tt ISAJET}, the relevant LSP decays must be
explicitly added (by hand) to the {\tt ISAJET} decay table.


\section{Gauge-Mediated Low-Energy Supersymmetry Breaking}

\subsection{Introduction}

Supersymmetry breaking must be transmitted from the
supersymmetry-breaking sector to the visible sector through
some messenger sector.
Most phenomenological studies of supersymmetry implicitly assume
that messenger-sector interactions are of gravitational strength.
It is possible, however, that the messenger scale for transmitting
supersymmetry breaking is anywhere between the Planck and just
above the electroweak scale.

The possibility of
supersymmetry breaking at a low scale has two important
consequences.
First, 
it is likely that
the standard-model gauge interactions play some role in the
messenger sector.
This is because standard-model gauginos couple
at the renormalizable level only through
gauge interactions.
If Higgs bosons received mass predominantly
from non-gauge interactions,
the standard-model gauginos would be unacceptably lighter than
the electroweak scale.
Second, the gravitino is naturally the lightest supersymmetric
particle (LSP).
The lightest standard-model superpartner
is the next to lightest supersymmetric particle (NLSP).
Decays of the NLSP to its partner plus the Goldstino component
of the gravitino within a detector lead to very distinctive
signatures.
In the following subsections the minimal model of gauge-mediated
supersymmetry breaking, and the experimental
signatures of decay to the Goldstino, are presented.

\subsection{The Minimal Model of Gauge-Mediated
Supersymmetry Breaking}

The standard-model gauge interactions act as messengers of
supersymmetry breaking if fields within the supersymmetry-breaking
sector transform under the standard-model gauge group.
Integrating out these messenger-sector fields gives rise to
standard-model gaugino masses at one-loop, and scalar masses
squared at two loops.
Below the messenger scale the particle content is just that of
the MSSM plus the essentially massless Goldstino discussed in
the next subsection.
The minimal model of gauge-mediated supersymmetry breaking
(which preserves the successful predictions of perturbative
unification) consists of messenger fields which transform as a single
flavor of ${\bf 5} + {\bar{\bf 5}}$ of $SU(5)$, i.e.  there are
triplets, $q$ and $\bar{q}$, and doublets, $\ell$ and $\bar{\ell}$.
These fields couple to a single gauge singlet field, $S$,
through the superpotential
\begin{equation}
W = \lambda_3 S q \bar{q} + \lambda_2 S \ell \bar{\ell}.
\end{equation}
A non-zero expectation value for the scalar component of
$S$ defines the messenger scale, $M = \lambda S$, while
a non-zero expectation value for the auxiliary component, $F$,
defines the supersymmetry-breaking scale within the messenger sector.
For ${F} \ll \lambda S^2 $,
the one-loop visible-sector gaugino masses at the messenger scale
are given by \cite{gms}
\begin{equation}
m_{\lambda_i}=c_i\ 
{\alpha_i\over4\pi}\ \Lambda\
\end{equation}
where $c_1 =c_2=c_3=1$ (we define $g_1=\sqrt{5\over 3}g'$), and
$\Lambda = F / S $.
The two-loop squark and slepton masses squared at the messenger
scale are \cite{gms}
\begin{equation}
\tilde{m}^2 ={2 \Lambda^2} 
\left[
C_3\left({\alpha_3 \over 4 \pi}\right)^2
+C_2\left({\alpha_2\over 4 \pi}\right)^2
+{3 \over 5}{\left(Y\over2\right)^2}
\left({\alpha_1\over 4 \pi}\right)^2\right]
\end{equation}
where $C_3 = {4 \over 3}$
for color triplets and zero for singlets, $C_2= {3 \over 4}$ for
weak doublets and zero for singlets,
and $Y$ is the ordinary hypercharge normalized
as $Q = T_3 + {1 \over 2} Y$.
The gaugino and scalar masses go roughly as their gauge
couplings squared.
The Bino and right-handed sleptons gain masses only through
$U(1)_Y$ interactions, and are therefore lightest.
The Winos and left-handed sleptons, transforming under $SU(2)_L$,
are somewhat heavier.
The strongly interacting squarks and gluino are significantly
heavier than the electroweak states.
Note that the parameter $\Lambda = F / S $ sets the scale for the
soft masses (independent of the $\lambda_i$ for ${F} \ll \lambda S^2$).
The messenger scale $M_i$, may be anywhere between roughly 100 TeV and
the GUT scale.

The dimensionful parameters within the Higgs sector,
$W = \mu H_u H_d$ and $V = m_{12}^2 H_u H_d + h.c.$,
do not follow from the ansatz of gauge-mediated supersymmetry breaking,
and require
additional interactions.
At present there is no good model which gives rise to these
Higgs-sector masses without tuning parameters.
The parameters $\mu$ and $m_{12}^2$ are therefore taken as free
parameters in the minimal model, and can be eliminated as usual in favor
of $\tan \beta$ and $m_Z$.

Electroweak symmetry breaking results (just as for high-scale
breaking) from the negative one-loop
correction to $m_{H_u}^2$ from stop-top loops
due to the large top quark Yukawa coupling.
Although this effect is formally three loops, it is larger in magnitude
than the electroweak contribution to $m_{H_u}^2$
due to the large squark masses.
Upon imposing electroweak symmetry breaking, $\mu$ is typically found to be
in the range $\mu \sim (1-2) m_{\tilde{\ell}_L}$ (depending on
$\tan \beta$ and the messenger scale).
This leads to a lightest neutralino, $\tilde{\chi}_1^0$, which is mostly
Bino, and a lightest chargino, $\tilde{\chi}_1^{\pm}$, which is mostly
Wino.
With electroweak symmetry breaking imposed,
the parameters of the minimal model may be taken to be
\begin{equation}
(~\tan \beta~,~ \Lambda = F/S~,~ {\rm sign}~\mu~,~ \ln M~)
\end{equation}
The most important parameter is $\Lambda$
which sets the overall scale for the superpartner spectrum.
It may be traded for a physical mass, such
as $m_{{\tilde{\chi}_1^0}}$ or $m_{\tilde{\ell}_L}$.
The low energy spectrum is only weakly sensitive to $\ln M_i$, and the
splitting between $\ln M_3$ and $\ln M_2$ may be neglected for most
applications.

\subsection{The Goldstino}

In the presence of supersymmetry breaking
the gravitino gains a mass by the super-Higgs mechanism
\begin{equation}
m_{ G}=\frac{F}{\sqrt{3}M_{p}} \simeq 2.4\, \left(
\frac{F}{(100~{\rm TeV})^2} \right) \rm{eV}
\end{equation}
where $M_p \simeq 2.4 \times 10^{18}$ GeV is the reduced
Planck mass.
With low-scale supersymmetry breaking the gravitino is naturally the
lightest supersymmetric particle.
The lowest-order couplings of the spin-${1 \over 2}$
longitudinal Goldstino component of the gravitino, $G_{\alpha}$,
are fixed by the supersymmetric Goldberger-Treiman low energy
theorem to be given by \cite{gdecay}
\begin{equation}
L=-\frac{1}{F}j^{\alpha\mu}\partial_{\mu}G_{\alpha}+h.c.
\label{goldcoupling}
\end{equation}
where $j^{\alpha\mu}$ is the supercurrent.
Since the Goldstino couplings (\ref{goldcoupling})
are suppressed compared to electroweak
and strong interactions, decay to the Goldstino is only relevant for the
lightest standard-model superpartner (NLSP).

With gauge-mediated supersymmetry breaking it is natural that
the NLSP is either a neutralino (as occurs in the minimal model)
or a right-handed slepton (as occurs for a messenger sector
with two flavors of ${\bf 5} + \bar{\bf 5}$).
A neutralino NLSP can decay by
$\tilde{\chi}_1^0 \rightarrow (\gamma, Z^0, h^0, H^0, A^0) + G$, while
a slepton NLSP decays by $\tilde{\ell} \rightarrow \ell + G$.
Such decays
of a superpartner to its partner plus the Goldstino
take place over a macroscopic distance, and for
$\sqrt{F}$ below a few 1000 TeV,
can take place within a detector.
The decay rates into the above
final states can be found in
Ref.~ \cite{DimopoulosPRL,DimopoulosSecond,Grav,kolda}.

\subsection{Experimental Signatures of Low-Scale Supersymmetry Breaking}

The decay of the lightest standard-model superpartner
to its partner plus the Goldstino within a detector
leads to very distinctive
signatures for low-scale supersymmetry breaking.
If such signatures were established experimentally, one of the
most important challenges would be to measure the distribution
of finite path lengths for the NLSP,
thereby giving a direct measure of the supersymmetry-breaking
scale.

\subsubsection{Neutralino NLSP}

In the minimal model of gauge-mediated supersymmetry breaking,
$\tilde{\chi}_1^0$ is the NLSP.
It is mostly gaugino and decays predominantly by
$\tilde{\chi}_1^0 \rightarrow \gamma + G$.
Assuming $R$ parity conservation, and decay within the detector,
the signature for supersymmetry at a collider is then
$\gamma \gamma X + {\not \! \! E_{T}}$,
where $X$ arises from cascade decays to $\tilde{\chi}_1^0$.
In the minimal model the strongly interacting states are
much too heavy to be relevant to discovery, and it is
the electroweak states which are produced.
At $e^+e^-$ colliders $\tilde{\chi}_1^0$ can
be probed directly by $t$-channel $\tilde e$ exchange, yielding
the signature $e^+e^-\to \tilde{\chi}^0_1\tilde{\chi}^0_1\to\gamma\gamma +
{\not \! \! E_{T}}$.
At a hadron collider the most promising signals include
$ q q' \rightarrow \tilde{\chi}^0_2\tilde{\chi}^\pm_1, \tilde{\chi}_1^+
\tilde{\chi}_1^- \rightarrow
 \gamma \gamma X + {\not \! \! E_{T}}$, where
$ X = WZ, WW, W\ell^+\ell^-,\dots$.
Another clean signature is $q q' \rightarrow
\tilde{\ell}^+_R \tilde{\ell}^-_R \rightarrow \ell^+\ell^-
\gamma\gamma+{\not \! \! E_{T}}$.
One event of this type has in fact been reported by the CDF
collaboration~ \cite{event}.
In all these signatures both the missing energy and photon energy
are typically greater than $m_{\tilde{\chi}^0_1}/2$.
The photons are also generally isolated.
The background from initial- and final-state
radiation typically has non-isolated photons with a much
softer spectrum.

In non-minimal models it is possible for $\tilde{\chi}_1^0$ to have
large Higgsino components, in which case $\tilde{\chi}_1^0 \rightarrow h^0+G$
can dominate.
In this case the signature $bbbbX + {\not \! \! E_{T}}$ arises
with the $b$-jets reconstructing $m_{h^0}$ in pairs.
This final state topology may be difficult to reconstruct
at the LHC -- a systematic study has not yet been attempted.

Detecting the finite path length associated with
$\tilde{\chi}_1^0$ decay represents a major experimental challenge.
For the case $\tilde{\chi}_1^0 \rightarrow \gamma + G$, tracking within
the electromagnetic calorimeter (EMC) is available.
A displaced photon
vertex can be detected as a non-zero impact parameter
with the interaction region.
For example, with a photon angular resolution of 40 mrad/$\sqrt{E}$
expected in the CMS detector with a preshower array covering
$| \eta | < 1$ \cite{CMS},
a sensitivity to
displaced photon vertices of about 12 mm at the 3$\sigma$ level results.
Decays well within the EMC or hadron calorimeter (HC)
would give a particularly distinctive signature.
In the case of decays to charged particles, such as from
$\tilde{\chi}_1^0 \rightarrow (h^0, Z^0) +G$ or
$\tilde{\chi}_1^0 \rightarrow \gamma^* +G$ with $\gamma^* \rightarrow f
\bar{f}$,
tracking within a silicon vertex detector (SVX) is available.
In this case displaced vertices down to the 100 $\mu$m level should
be accessible.
In addition, decays outside the SVX, but inside the EMC, would
give spectacular signatures.

\subsubsection{Slepton NLSP}

It is possible within non-minimal models that a right-handed slepton
is the NLSP, which decays by $\tilde{\ell}_R \rightarrow \ell + G$.
In this case the signature for supersymmetry
is $\ell^+\ell^- X + {\not \! \! E_{T}}$.
At $e^+e^-$ colliders such signatures are fairly clean.
At hadron colliders
some of these signatures have backgrounds from $WW$ and $t \bar{t}$
production.
However, $\tilde{\ell}_L \tilde{\ell}_L$ production can give $X=4\ell$,
which has significantly reduced backgrounds.
In the case of $\tilde{\ell}_R \tilde{\ell}_R$ production
the signature is nearly identical to slepton pair production with
$\tilde{\ell} \rightarrow \ell + \tilde{\chi}_1^0$ with
$\tilde{\chi}_1^0$ stable.
The main difference here is that the missing energy is carried
by the massless Goldstino.

The decay $\tilde{\ell} \rightarrow \ell + G$ over a macroscopic distance
would give rise to the spectacular signature of
a greater than minimum ionizing track
with a kink to a minimum ionizing track.
Note that if the decay takes place well outside the
detector, the signature for supersymmetry is heavy charged particles
rather than the traditional missing energy.

\subsection{Event Generation}

For event generation by {\tt ISAJET}, the user must provide a program to
generate the appropriate spectra for a given point in the above
parameter space.  The corresponding $MSSMi$ parameters can be entered into
{\tt ISAJET} to generate the decay table, except for the NLSP decays to
the Goldstino.  If $NLSP\rightarrow G+\gamma$ at 100\%, the $FORCE$ command can
be used.  Since the $G$ particle is not currently defined in {\tt ISAJET},
the same effect can be obtained by forcing the NLSP to decay to a neutrino
plus a photon.  If several decays of the NLSP are relevant, then each decay
along with its branching fraction must be explicitly added to the {\tt ISAJET}
decay table.  Decay vertex information is not saved in {\tt ISAJET}, so that
the user must provide such information.
In {\tt Spythia}, the $G$ particle is defined, and decay vertex
information is stored.

\section{Conclusions}

In this report we have looked beyond the discovery
of supersymmetry, to the even more exciting prospect of
probing the new physics (of as yet unknown type) which
we know must be associated with supersymmetry and
supersymmetry breaking.
The collider experiments which
disentangle one weak-scale SUSY scenario from another will
also be testing hypotheses about new physics at very high
energies: the SUSY-breaking scale, intermediate symmetry-breaking
scales, the GUT scale, and the Planck scale.

We have briefly surveyed the variety of ways that
weak-scale supersymmetry may manifest itself at colliding
beam experiments.
We have indicated for each SUSY scenario how Monte Carlo
simulations can be performed using existing event generators
or soon-to-appear upgrades.  In most cases very little
simulation work has yet been undertaken.  Even in the case
of minimal supergravity the simulation studies to date
have mostly focused on discovery reach, rather than the
broader questions of parameter fitting and testing key
theoretical assumptions such as universality.
Clearly more studies are needed.

We have seen that alternatives to the
minimal supergravity scenario often provide distinct
experimental signatures.  Many of these signatures involve
displaced vertices: the various NLSP decays, LSP decays
from $R$ parity violation, chargino decays in the 200 and O-II
models, and enhanced $b$ multiplicity in the 24 model.
This observation emphasizes the crucial importance of
accurate and robust tracking capabilities in future collider
experiments.

The phenomenology of some scenarios is less dramatic and thus
harder to distinguish from the bulk of the mSUGRA parameter
space.  In any event, precision measurements will be
needed in the maximum possible number of channels.
In the absence of a ``smoking gun'' signature like those
mentioned above, the most straightforward way to identify
variant SUSY scenarios will be to perform an overconstrained
fit to the mSUGRA parameters.  Any clear inconsistencies in
the fit should point to appropriate alternative scenarios.
More study is needed of how to implement this procedure in
future experiments with real-world detectors and data.


\end{document}